# The impact of the ISR on accelerator physics and technology

*P.J. Bryant*

**Abstract**

The ISR (Intersecting Storage Rings) were two intersecting proton synchrotron rings each with a circumference of 942 m and eight-fold symmetry that were operational for 13 years from 1971 to 1984. The CERN PS injected 26 GeV/*c* proton beams into the two rings that could accelerate up to 31.4 GeV/*c*. The ISR worked for physics with beams of 30–40 A over 40–60 hours with luminosities in its superconducting low-β insertion of $10^{31}$–$10^{32}$ cm$^{-2}$ s$^{-1}$. The ISR demonstrated the practicality of collider beam physics while catalysing a rapid advance in accelerator technologies and techniques.

## Introduction

To appreciate the role played by the CERN Intersecting Storage Rings (ISR) in accelerator physics let us try to imagine ourselves back in the 1960s. In an atmosphere of close collaboration and friendly competition, Europe and America had just commissioned the first generation of strong-focusing machines, the CERN PS (1959) and the BNL AGS (1960). With the PS working, CERN embarked on the study of a proton–proton collider leading to the approval of the ISR in 1965. The ISR [1] was an exciting concept that offered a giant leap in the centre-of-mass energy over the fixed-target configuration, but it was clouded by doubts as to its practicality, and support was far from unanimous. On the one hand, the CERN team was highly experienced by 1960 standards and well connected to the other leading laboratories, but on the other hand there were voices saying that the residual gas and non-linear resonances would destroy the beams, since there was no stabilizing influence from synchrotron radiation. These were not empty fears. The vacuum problem was very real and there is indeed an infinite web of non-linear resonances [2], [3]. The deleterious effect of electrons trapped in the potential well of the beam following ionization of the residual gas was also foreseen by the early designers [4]. With hindsight we know these effects were not going to be fatal, but could one be sure in 1965? We also know that the CERN team was standing at the foot of a steep and exciting learning curve in accelerator technology, techniques, and diagnostics. By the time the ISR closes in 1984, our concepts of accelerator engineering and diagnostics and the way experimental physics is conducted will have changed radically and will look very much as they do today. The aim of this paper is to illustrate the rapid change that took place, mainly in the years 1965 to 1977, and to underline the role played by the ISR.

## 1    Advances in lattice design

*The '1965' lattice*

Figure 1 shows the layout of the two rings with their injection lines and Fig. 2 shows the original ISR lattice functions from the centre of an outer arc to the centre of an inner arc. The underlying lattice is a FFDD structure, but in the crossing regions and inner arcs the lattice is opened between the F units to form FDDF cells. The longer drift spaces are a welcome innovation compared to the tightly-packed FD–DF cells of the PS. The ISR magnets are, however, still combined-function of the open 'C' type (see Fig. 3), a legacy from the days of weak focusing. Note that the betatron amplitude functions are very rounded. This is due to the spread-out gradients of the combined-function magnets. The lattice has been manipulated globally to fit the interlaced geometry and to provide space for physics equipment in the interaction regions. The 'split-F' structure provides local betatron minima at the crossing points, although these are not as low as would have been liked. There are no dispersion-free regions or low-β insertions as the local customization of a lattice (i.e. insertions) had still to be developed.

**Fig. 1:** Layout of the CERN ISR with transfer lines (Design Study 1964)

**Fig. 2:** Design lattice functions of the ISR (based on ISR Parameter List Rev. 5 CERN/ISR-GS/76-4)

**Fig. 3:** ISR main magnet model (December 1965)

*The '1977' SCISR upgrade lattice*

Some years later in 1977, a project was published to convert the ISR to the SCISR, a superconducting machine [5]. Figure 4 shows the lattice functions in the outer arc for this conversion and Fig. 5 shows the cross-section of the new quadrupole. Immediately, one sees the advances that have taken place. The arc is a tightly-packed FODO structure, terminated by a dispersion suppressor and matched into a low-$\beta$ insertion, which is exactly how the job would be done today. The use of separated-function magnets provides a more efficient focusing with clearer features in the shapes of the lattice functions, and the superconducting quadrupole using the Roman arch principle to support the coils is right up to date. This is just one illustration of the rapid changes mentioned in the introduction that were to take place during the brief 13-year life of the ISR. Eberhard Keil [6] had proposed the elegant method used for the dispersion suppressor and, although low-$\beta$ insertions were not an ISR invention, the matching was based on an analytical solution for a variable-geometry triplet published by Bruno Zotter [7] in 1973. This is perhaps the most useful of all the analytical matching modules ever published.

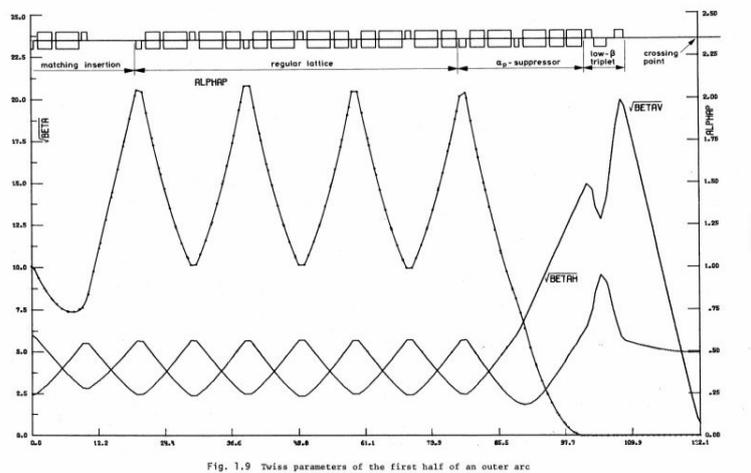

**Fig. 4:** Lattice functions of the proposed superconducting ISR upgrade (1977)

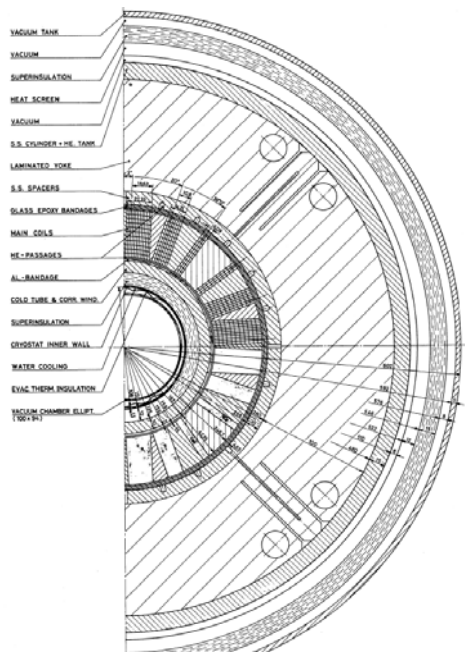

**Fig. 5:** The proposed superconducting ISR quadrupole upgrade (1977)

## 2 From 'global fitting' to insertions

*Terwilliger scheme*

One example of 'global fitting' with the 1965 lattice is the so-called Terwilliger Scheme [8], which creates small interaction diamonds by driving the dispersion function to zero at regularly-spaced positions in betatron phase using a superimposed gradient with a suitable azimuthal harmonic. In the ISR, which is the only machine to have demonstrated this now obsolete principle [9], only four of the eight minima fell on interaction regions (see Fig. 6).

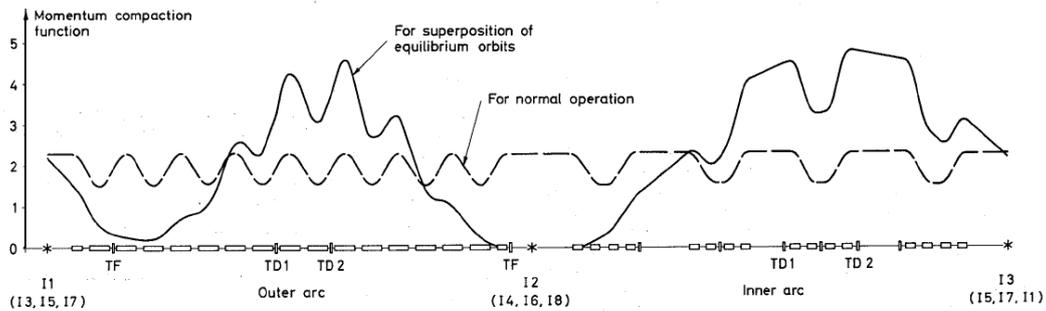

**Fig. 6:** The unperturbed and perturbed momentum compaction functions through one superperiod of the ISR showing how the small interaction diamonds are formed (1973)

*Conventional steel low-β insertion*

The concept of 'global fitting' is to be compared to the more modern idea of 'insertions' that tailor the lattice locally for a particular task. By 1974, the concept of a local insertion had been demonstrated in the ISR by a conventional steel low-β insertion built in Intersection 7 using largely borrowed quadrupoles from the CERN PS, DESY, and the Rutherford Appleton Laboratory. Since the ISR had no dispersion-free regions and the lattice functions were far from regular, matching the low-β was a significant challenge, but the result was highly successful and increased the luminosity by a factor of 2.3 [10]. The steel low-β insertion was initially an 'experiment' to test the fear that the very marked super-periodicity of unity would cause the high-intensity ISR beams to be unstable or noisy. In reality, this did not prove to be an issue and two years later in 1976, the insertion was demounted and moved to Intersection 1, where it was used in conjunction with a superconducting solenoid, see Fig. 7. It remained operational until the closure of colliding beams.

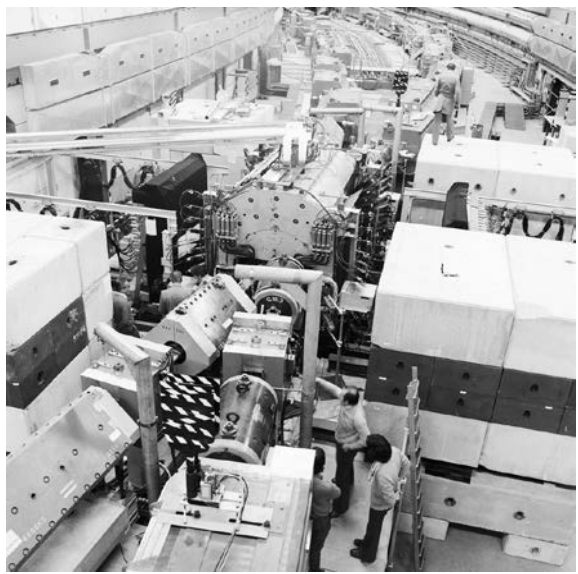

**Fig. 7:** Steel low-β insertion in Intersection 1 (1976)

## 3  Lattice programs

During the construction of the ISR, lattice computations were made with the programs SYNCH [11] (LBL), AGS [12] and BEATCH [13] (CERN). By the time the ISR closed in 1984, the ISR Theory Group had replaced the CERN AGS code by MAD (Methodical Accelerator Design) [14], which is now a de facto world standard for the study and design of large synchrotrons like LHC. Lattice programs are the essential tools behind lattice design and beam simulations. The effort devoted to these tools during the ISR years was very important to CERN, since CERN now holds the 'gold standard' software for one of the core competences of accelerator building. Similarly, at the start of the ISR, Romeo Perin, Simon van der Meer and Steve Caeymaex were working on computer codes for 2D [15] and 3D [16] magnet design, but these topics were not carried to the same level.

## 4  Coupling

The ISR also contributed strongly to the theory and design of coupling compensation schemes. A complete Hamiltonian theory for sum and difference resonances was published in 1976 by Gilbert Guignard [17], in which, amongst other things, the driving terms and coupling coefficients are defined. It later turned out that Phil Morton from SLAC had reached many of the same results in an unpublished and unfinished note [18]. A story that is similar to those of Rolf Wideröe and his betatron and Lee Teng who did not publish his theory for the rotator for medical gantries. As the operation of the ISR progressed, there were practical applications of the theoretical work. Since there were no dispersion-free regions in the ISR, the compensation scheme for the global coupling was of a special and unique design [19]. Similarly, the physics solenoid in Intersection 1 had horizontal slots in its end plates to accommodate the beams that crossed at an angle. This new feature was described analytically and compensated. The ISR was also first to be equipped with an electronic coupling meter that directly gave the modulus of the coupling coefficient defined in the theory [20].

## 5  Advances in magnet technology

### Poleface windings

Figure 3 shows the ISR model magnet that clearly has a close affinity to the CERN PS magnet. Although the combined-function and C-type construction of the ISR main magnet was not according to modern tastes, it did have an extremely versatile set of poleface windings set into a thick epoxy cover (Fig. 8) placed over the pole under a copper heat shield[*]. The F-blocks and D-blocks were each equipped with 12 circuits and a 13th circuit in each case to compensate the stray field from the cable bundle. Probably no other machine has ever had such complete control over higher-order field components. The field shaping was applied using a 'practical' system of so-called 'half-multipoles' that acted independently on the inner and outer halves of the aperture. I will return to the use of this system under space-charge loading corrections in the section on beam–chamber interactions.

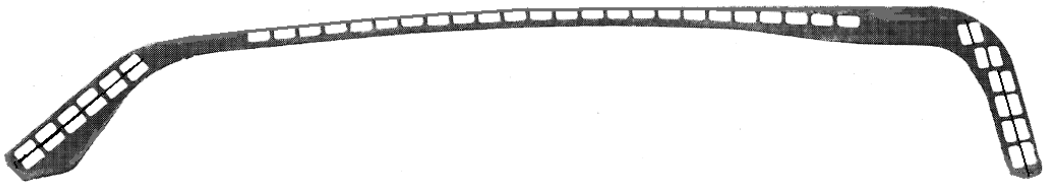

**Fig. 8:** Cross-section of the poleface windings sheath that was mounted on each pole of the ISR main magnet under the heat shield

---

[*] Eddy currents in the heat shield played a significant role in the stability of the fields against ripple when coasting.

*Superconducting quadrupoles for a low-β insertion*

It has already been mentioned that a superconducting upgrade for ISR was published in 1977. This upgrade was not approved, but another project to build a superconducting low-β insertion [21] had already been accepted. At that time, CERN made an important decision not to outsource the work to a Member State laboratory such as Rutherford, UK, but rather to start accumulating in-house expertise in magnet building and cryogenics, which was later to be immensely important for LEP and LHC. A number of superconducting magnets already existed in transfer lines in various laboratories, but nobody had operated superconducting quadrupoles in the lattice of a synchrotron. CERN then made a second important decision to build the models and prototypes in-house in order to reduce the new technology to a detailed engineering specification. On the basis of this specification, tenders were then invited from industry for the series production. This was a middle-of-the-road approach between the extremes of building everything in-house, as is usually done in US laboratories, or requesting industry to do the R&D as well as the series production, as some Member States wanted CERN to do. In the ISR approach, it is made clear to the manufacturer that the magnetic design is the responsibility of CERN and that he, the manufacturer, is only responsible for respecting the tolerances, choices of materials, and adherence to the various qualified procedures. The superconducting low-β, see Fig. 9, was a great success, increasing the luminosity in Intersection 8 by a factor of 6.5 [22] and laying the foundations for the LHC magnets and cryogenics. This was the first time that industrially-built superconducting quadrupoles had been operated in the lattice of a synchrotron for regular operation. In a typical run, the magnets would operate at top energy for 60 hours.

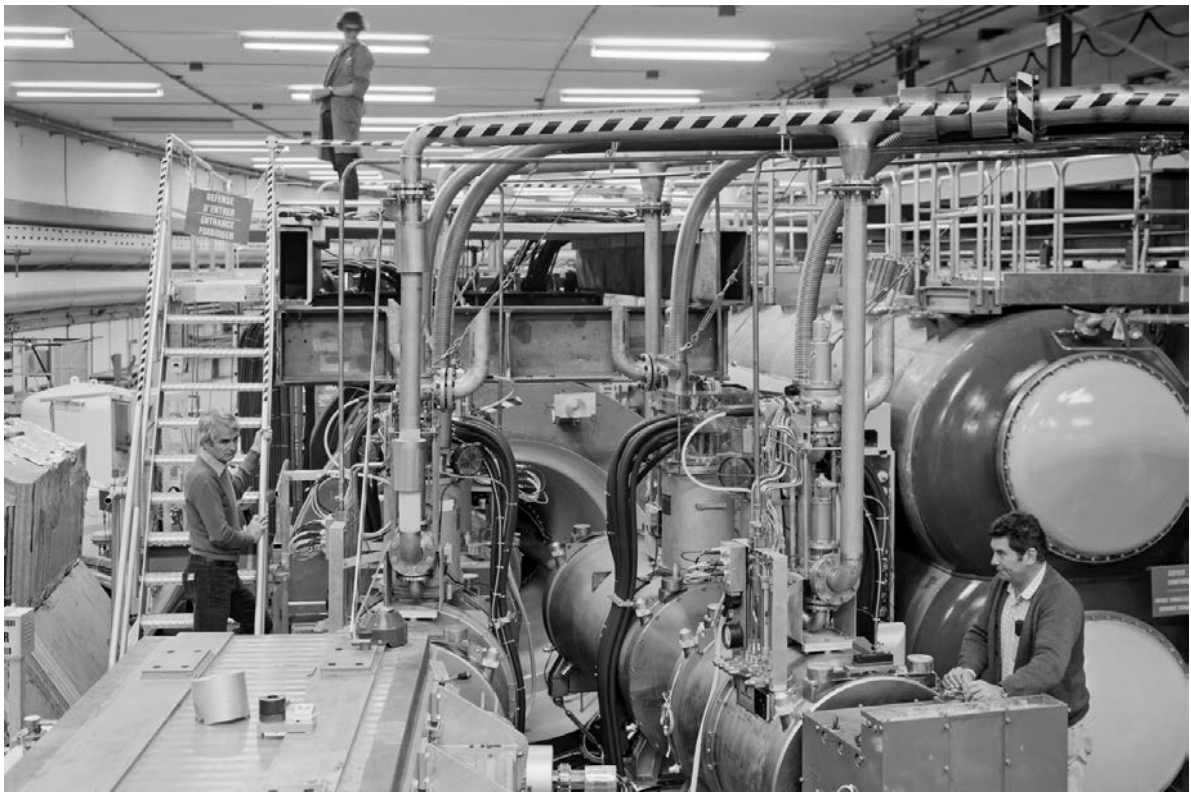

**Fig. 9:** Superconducting low-β insertion in Intersection 8 (1980)

At the start of a physics run in December 1982, the record luminosity of $1.4 \times 10^{32}$ cm$^{-2}$ s$^{-1}$ was measured in the superconducting low-β insertion. This record was not beaten until 1991 by CESR at Cornell with $1.7 \times 10^{32}$ cm$^{-2}$ s$^{-1}$. Figure 10 shows the history of the maximum luminosity in the ISR over its working life. It is interesting to note that the design luminosity of $4 \times 10^{30}$ cm$^{-2}$ s$^{-1}$ was reached within the first 2 years and thereafter rose by nearly 2 orders of magnitude.

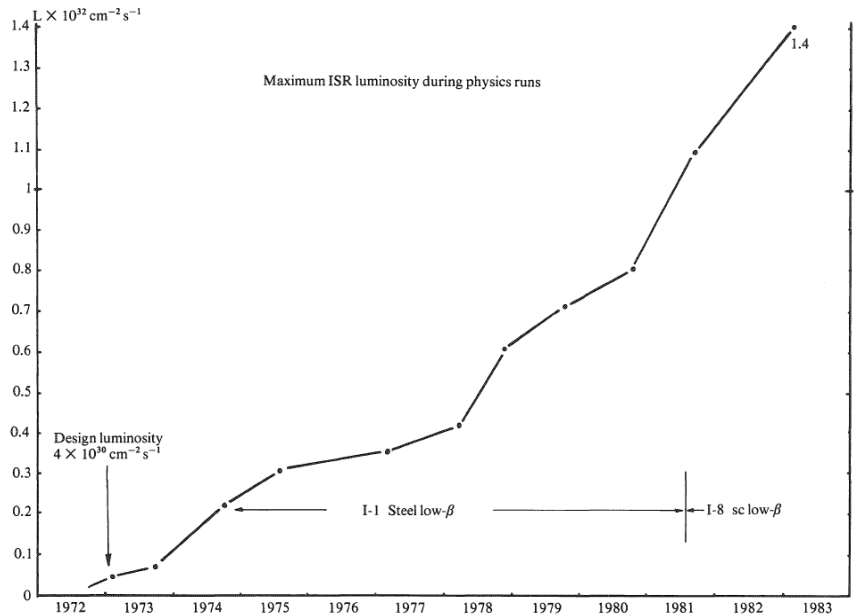

**Fig. 10:** History of the maximum luminosity in the ISR

*Physics detector magnets*

In 1965, colliding beam physics was more or less a blank page. In the official history of CERN, it was said that ISR was regarded by some as "an expensive small-angle scattering experiment for the PS". By the late 1970s, detector magnets with $4\pi$ acceptance were standard equipment for the ISR. The Superconducting Solenoid was installed in Intersection 1, the Open Axial Field Magnet was installed in Intersection 8 and an Air-cored Toroid in Intersection 6. This was effectively a single jump to today's technologies in just a few years. The principles had been established and only the scale of the equipment would increase further. The Open Axial Field Magnet (OAFM) is shown in Fig. 11 because it is perhaps the least well known of the examples.

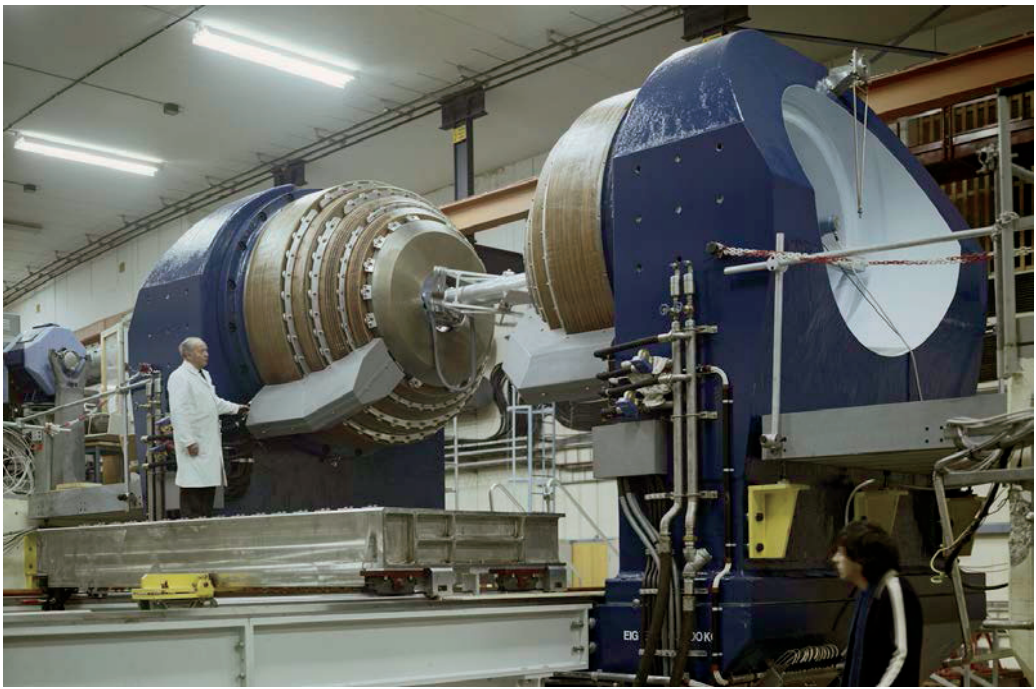

**Fig. 11:** The Open Axial Field Magnet in Intersection 8 (1979)

# 6  Vacuum

*Base vacuum*

The original design criterion stated that the lifetime imposed by gas scattering should be at least one order of magnitude longer than the time needed to fill the rings. This was interpreted as $10^{-9}$ torr in the arcs (using ion pumps) and $10^{-10}$ torr in the crossing regions (using cryo-pumps), although it was noted that $10^{-11}$ torr would be more desirable. The chamber itself was to be stainless steel bakeable to 300°C to 350°C, although initially it was only baked to 200°C. This made the ISR the world's largest ultra-high vacuum (UHV) system, which was an enormous challenge for the technology of the time. In fact, the pressure in the arcs was $10^{-10}$ torr (beams < 2 A) for the first run in January 1971, and $10^{-11}$ torr (beams < 2 A) was quickly reached in the crossing regions in 1972. In subsequent years, the pressure fell further to an average around the machine close to $10^{-12}$ torr. Figure 12 shows the evolution of the average pressure over the lifetime of the machine. This improvement meant that the principal source of background in the experiments became the beam losses from non-linear resonances. This led to hundreds of hours of machine development investigating working lines in the tune diagram, halo scraping exercises (called beam cleaning) and tests with cooling.

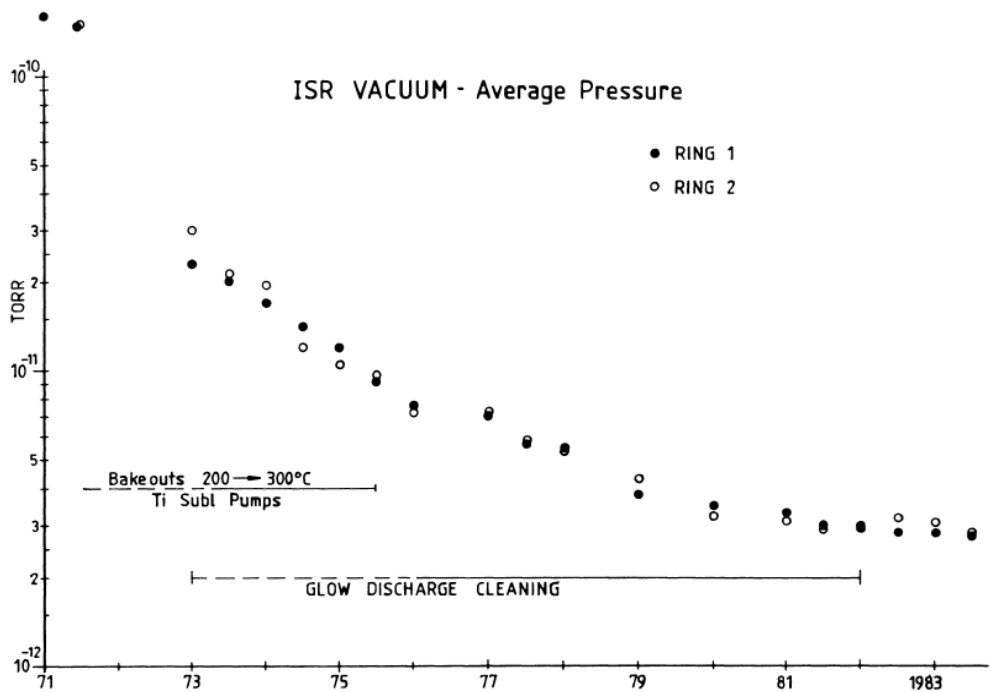

**Fig. 12:** Evolution of the average pressure in the ISR over its lifetime

*Beam-induced vacuum instability*

It was quickly discovered, however, that the beam and the vacuum could mutually destroy themselves in more ways than one. As the beam current increased, the residual gas was ionized and positive ions were repelled by the beam (the beam potential was typically 1 kV–2 kV) to crash into the chamber walls only to release adhered gas molecules that were in turn ionized by the beam to create a runaway effect that caused catastrophic beam loss. A staged programme to improve the vacuum system was started in 1971 and progressed over several years. Baking at 300°C and later at 350°C, instead of the initial 200°C helped and some 500 additional titanium sublimation pumps were added. The vacuum system was demounted arc by arc during shutdowns and cleaned using a new technique called glow-discharge cleaning. Later this was done in situ during bakeout by using the clearing electrodes to excite an argon discharge. Incredibly, a glow-discharged chamber could be opened to the air and left for many hours and still recover its ultra clean condition when pumped. This was the first demonstration of the efficiency of glow-discharge cleaning on a large scale [23].

*Neutralization from ionization of residual gas*

The unwanted neutralization caused by electrons trapped in the potential well of the beam following ionization of the residual gas was already mentioned in the introduction. The trapped charge would cause tune shifts and eventually beam instability. Bunched beams could, under the right circumstances, flush out such regions, but the only sure way was to install clearing electrodes. Although the ISR design had foreseen hundreds of these, the inadequacy of the clearing system was already evident in 1971 and many more electrodes had to be added. When the ISR became operational, it provided the perfect test bed for measurements and many papers were published on neutralization tune shifts, e–p instabilities, electron removal by RF clearing, and ion clearing in antiproton beams. A good pedagogic account is to be found in Ref. [24] which contains a large number of references to the ISR.

*Vacuum system design philosophy*

The ISR years saw another marked change in philosophy concerning vacuum systems. The '1965' design was based on ease of access with 'C' shaped magnets and bolted flanges for all sections of the vacuum system. When the TT6 transfer line was installed in the early 1980s to bring antiprotons into the ISR, the vacuum system was practically all welded. Changing a magnet meant cutting the chamber and re-welding; a philosophy that has been largely adopted by the LHC.

*Thin-walled chambers*

Another pioneering activity in the ISR was the design and production of large, thin-walled vacuum chambers for the intersection regions. Typically the chamber walls were 0.28 mm to 0.4 mm thick and the materials used were stainless steel and titanium. Figure 13 shows the example of a thin-walled chamber being installed in Intersection 7 in 1974. Beryllium was considered, but never used in the ISR; LHC has beryllium chambers in the physics intersections.

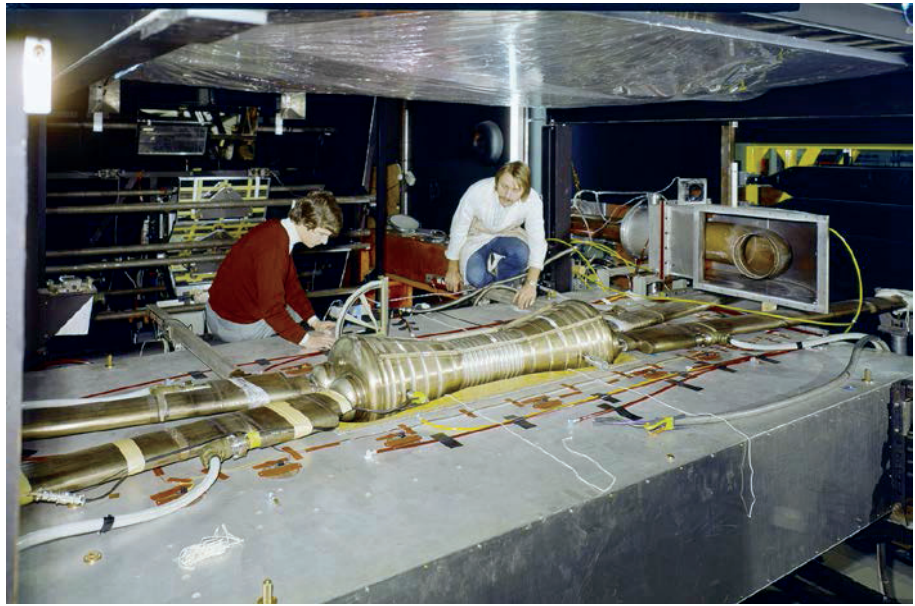

**Fig. 13:** Installing a thin-walled vacuum chamber in Intersection 7 (January 1974)

*Vacuum accidents*

By 1973 the ISR had suffered two catastrophic events caused by the beam burning holes in the vacuum chamber, see Fig. 14. Bellows were particularly vulnerable because each convolution would radiate onto its neighbour so the heat could only escape by conduction through the thin metal. This led to collimation rings being inserted in the flanges to protect the adjacent bellows. The thin-walled intersection chambers were also vulnerable to mechanical accidents as they were designed with small safety margins. The occasional collapse of such a chamber would leave a twisted sculpture and weeks of work to clean the contaminated arcs, see Fig. 15.

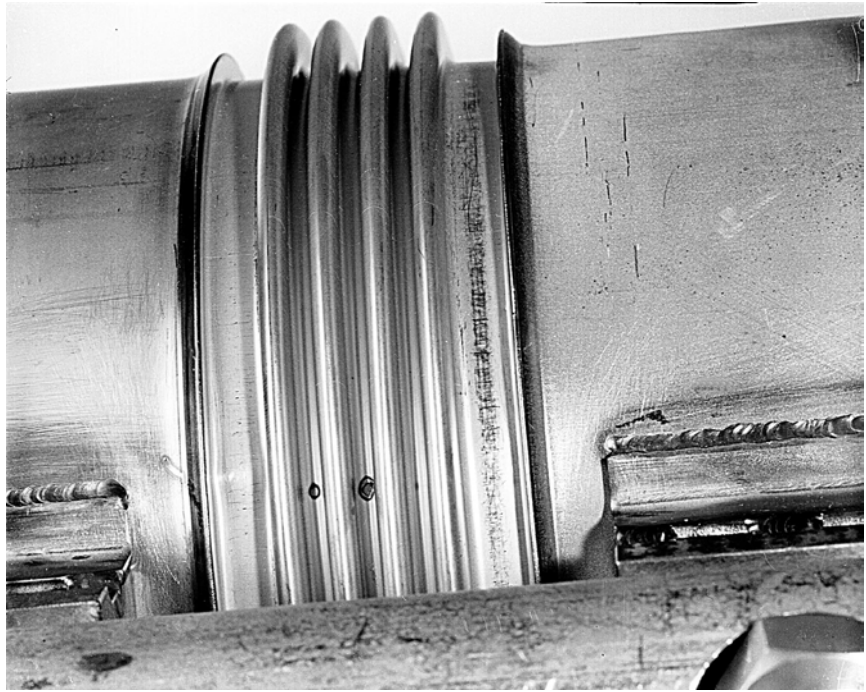

**Fig. 14:** Holes burnt by the beam in a bellows

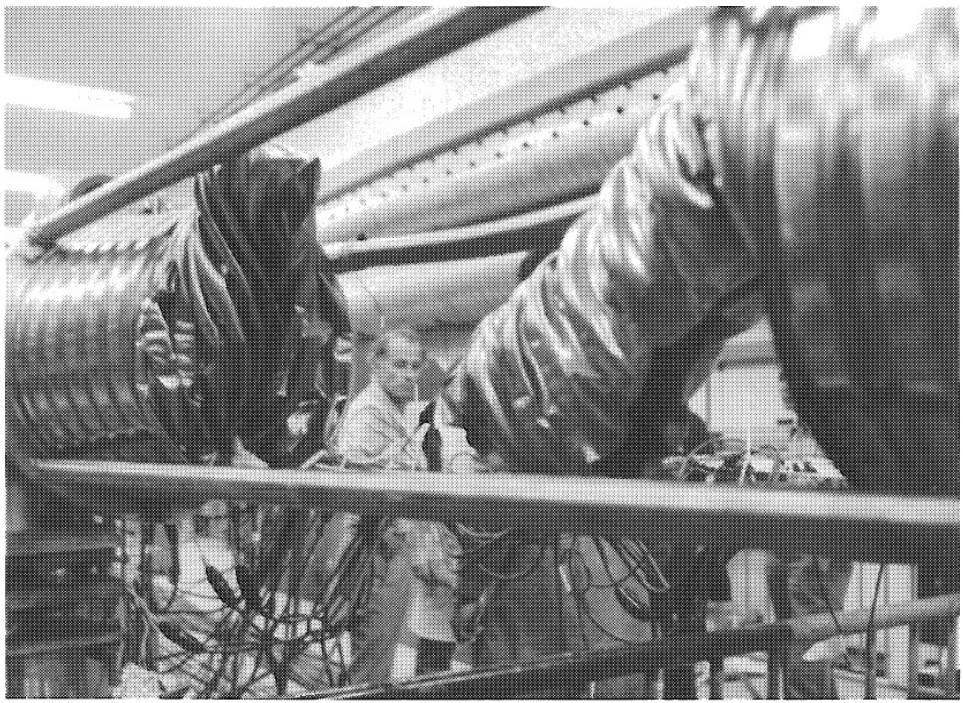

**Fig. 15:** The thin-walled (0.3 mm) titanium chamber in Intersection 7 implodes (1975)

# 7  Operation

*Computer control*

In 1965, the new dual Ferranti-Argus computers with a 16k core store of 24-bit words and a 1 µs store cycle time to a 640k disk store was not fully trusted and the control room was equipped with manual control panels some of which were physically locked by key to prevent unauthorized access. By the mid-term of the ISR attitudes had completely changed and a highly sophisticated control system was in place. Manual interventions were discouraged and automated procedures dominated physics operation, and for machine development high-level functions were used to control the machine.

*Closed-orbit correction*

One of the first operations to be carried out in any machine is the correction of the closed orbit. Two methods were developed for the ISR, out of which the algorithm MICADO [25] written by Bruno Autin was by far the more practical and efficient. MICADO was written into the COCO program [26] and after some years of development this became a de facto standard for many laboratories. The most advanced versions now have feedback loops and can be found in synchrotron light machines. A subject closely related to closed-orbit correction is injection optimization. After some years of development the ISR also had a sophisticated automatic injection procedure [27].

*Luminosity calibration method*

Luminosity calibration runs for the physics experiments were run as a semi-automatic procedure under computer control (LUMS program). The measurement method, which is still used today in the LHC, was invented in 1968 by Simon van der Meer, especially for the ISR colliding beam configuration [28]. The two beams are moved relative to each other in the vertical plane so that one beam effectively sweeps through the other while the interaction rate is monitored. The closed-orbit bumps used to move the beams were corrected for the coherent beam tune shift [29] and hysteresis in the so-called radial field magnets [30].

*Stacking*

Stacking in momentum space was an essential technique for accumulating the beam intensities needed to get a useful luminosity. In this scheme, proposed by MURA and tested in CESAR, the beam from the PS was slightly accelerated by the RF system in the ISR and the first pulse deposited at the highest acceptable momentum on an outer orbit in the relatively wide vacuum chamber. Subsequent pulses were added until the vacuum chamber was filled up to an orbit close to the injection orbit, which was on the inside of the chamber.

*Phase displacement acceleration*

Much of the operating life of the ISR was at 31.4 GeV/$c$, the maximum energy the magnet system could reach. After stacking at 26 GeV/$c$ the coasting beams were accelerated by the novel method of phase-displacement acceleration first suggested by MURA, but first extensively used in the ISR. The technique consisted of moving empty buckets repeatedly through the stacked beam from high to low energy. In accordance with longitudinal phase-space conservation, the whole stack was accelerated while the magnet field was simultaneously increased to keep the stack centred in the vacuum chamber. It was necessary to make many hundreds of sweeps with the empty buckets, which required a low-noise RF system operating at a low voltage with a fine control of the high-stability, magnet power supplies to prevent excessive beam blow-up. One amusing feature of this system was that the current would increase slightly as a beam was accelerated, a testimony to the extremely low losses that occurred during this operation.

*Other particles*

The ISR was also able to store deuterons and alpha particles as soon as they became available from the PS, leading to a number of runs with p–d, d–d, p–α and α–α collisions from 1976 onwards. For CERN's antiproton programme, a new beam line (TT6) was built from the PS to Ring 2 for antiproton injection. The first p–pbar runs took place in 1981. The ISR's final runs in 1984 were dedicated to a 3.5 GeV/$c$ antiproton beam colliding with a gas-jet target.

## 8 Diagnostics

*Schottky noise and stochastic damping*

If just one key discovery in the field of accelerator physics has to be singled out, then it would be Schottky noise — a statistical signal generated by the finite number of randomly distributed particles in a beam — which is well known to designers of electronic tubes. Simon van der Meer had worked on stochastic damping in 1968, but his ideas had seemed too far-fetched at that time. This changed in

1972 when Wolfgang Schnell actively looked for and found the longitudinal and transverse Schottky signals at the ISR [31], see Fig. 16. This prompted van der Meer to publish his work from 1968 about Schottky noise and the possibility of stochastic damping [32], opening new vistas for non-invasive beam diagnostics and active cooling systems for reducing the size and momentum-spread of a beam.

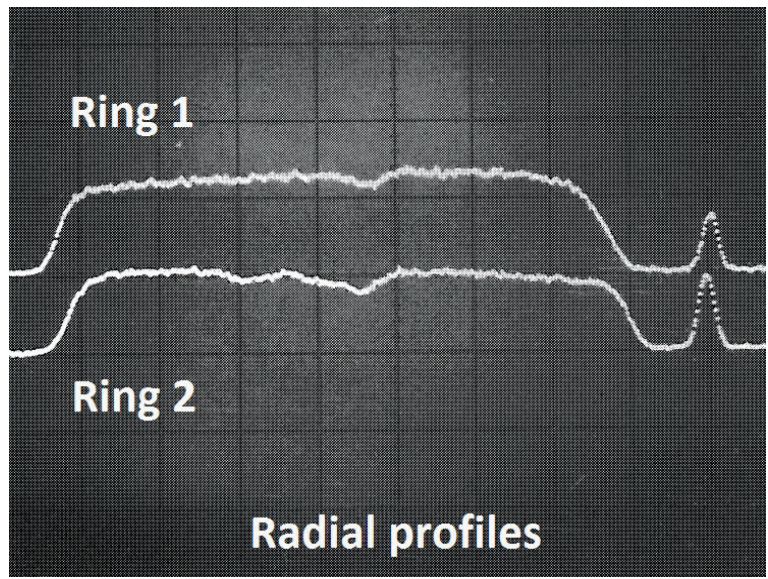

**Fig. 16:** First longitudinal Schottky results from the ISR (CERN Annual Report 1972, p. 109)

The longitudinal Schottky signal made it possible to measure the current density in the stack, without perturbing it, as a function of the momentum (transverse position), while the transverse Schottky signals gave information about how the density of the stack varied with the betatron frequency, or 'tune'. The combination of the two types of scan yielded a complete picture of the beam in the tune diagram, see Fig. 17, and the current density through the stack. These scans clearly show the beam edges and any markers. A marker could be created by using phase-displacement to accelerate part of the stack to create a narrow region of low current density, or by losses on resonances.

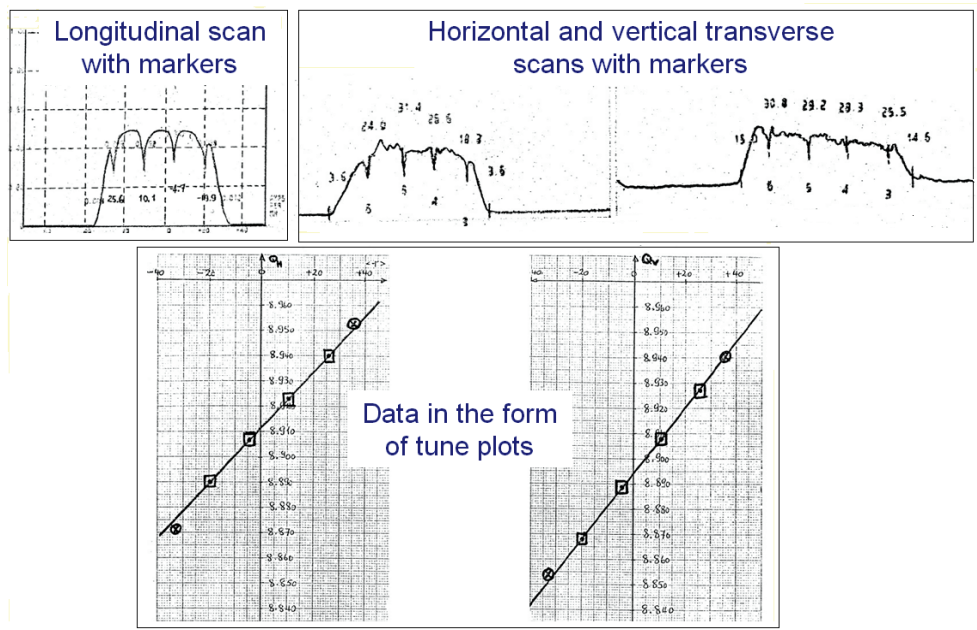

**Fig. 17:** Longitudinal and transverse Schottky scans used for beam diagnostics (ISR Performance report, S. Myers, 1977)

*Stochastic cooling (damping)*

The possibility of damping the betatron oscillations was experimentally demonstrated in the ISR in 1974 [33], see Fig. 18. Towards the end of the ISR's life, stochastic cooling was routinely used to cool antiproton beams in order to increase the luminosity in antiproton–proton collisions by counteracting the gradual blow-up of the antiproton beam through scattering with residual gas as well as resonances. Furthermore, stochastic cooling was the decisive factor in the conversion of the SPS to a p–pbar collider and hence in the discovery in 1983 of the long-sought-after W and Z bosons. Simon van der Meer and Carlo Rubbia shared the Nobel Prize in Physics in 1984.

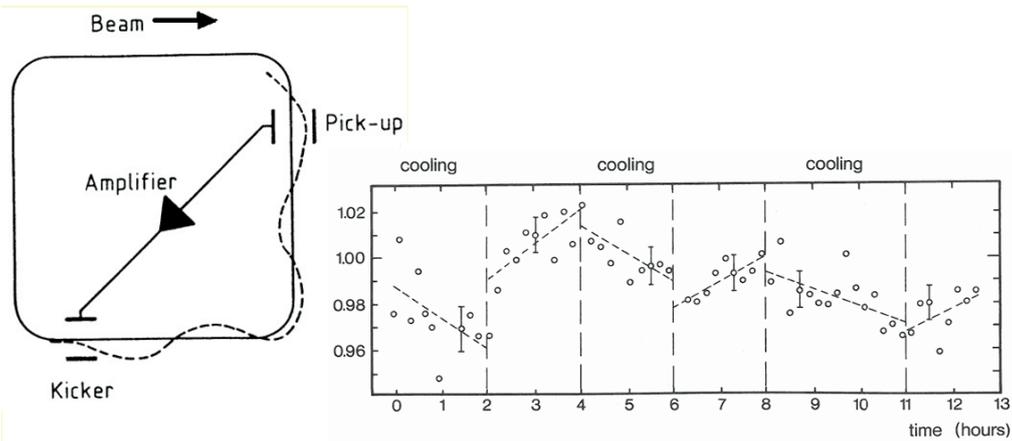

**Fig. 18:** Early results showing cooling in the ISR (CERN Annual Report 1974, p. 97)

Stochastic cooling became the cornerstone for the success, not only of the p–pbar collider in the SPS, but also for the more powerful Tevatron at Fermilab. CERN's low-energy antiproton programmes in the Low Energy Antiproton Ring and the Antiproton Decelerator, as well as similar programmes at GSI in Germany and at Brookhaven in the US, also owe their existence to stochastic cooling. The extension to bunched beams and to optical frequencies makes stochastic cooling today a basic accelerator technology.

*Direct-current beam transformer*

The ISR was also the home of the zero-flux, direct-current transformer (DCCT) [34], see Fig. 19. This device became another de facto world standard. Beam current monitors of this type developed at CERN in 1981 and 1990 became national primary standards in Germany, certified and operated by the PTB (Physikalisch-Technische Bundesanstalt) in Berlin.

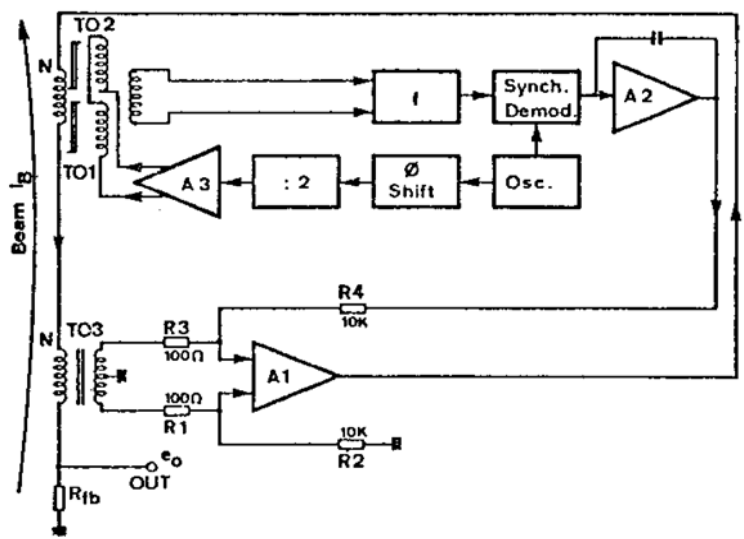

**Fig. 19:** Original zero-flux DCCT

# 9 Beam–chamber interaction

*'Brick wall' instability*

Shortly after the start of the ISR a coherent transverse instability was observed that limited the beam intensity. This phenomenon was dubbed the 'brick wall' instability [35], Fig. 20. The intensity would increase while stacking to around 3 A where there would be a sudden loss of 10% to 15% of the beam after which stacking would resume only to have a repeat beam loss at around the 3 A level. This gave a sawtooth pattern on the accumulated current that appeared to be knocking against a 'brick wall'. The phenomenon was due to the resistive wall instability [36] and it had been correctly predicted that a tune spread of 2 would stabilise the beams [37]. The additional tune spread was applied and the effect disappeared. Studies of the shape of the working line in the tune diagram revealed how the action of tune shifts induced by space-charge loading could destroy the tune spread and hence the stability of the beam (see insert in Fig. 20). From these beginnings, the ISR spawned tens of papers on incoherent and coherent space-charge tune shifts on central and off-axis orbits in variously shaped chambers and the correction of these effects.

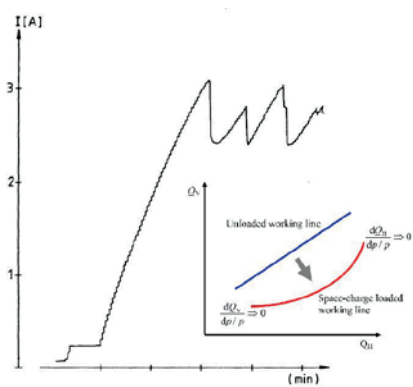

**Fig. 20:** Brick wall instability

*Pre-calculated and on-line space-charge corrections*

The first space-charge corrections to the working line in the tune diagram used the nominal beam density in momentum space to build a set of 'pre-stressed' working lines that when loaded with space charge would have the ideal shape and tune spreads for stability [38], see Fig. 21, but once longitudinal Schottky scans became operational the true current density could be measured at any time, the tune shifts with radial position calculated and the necessary poleface winding currents calculated and applied. Typically, these corrections would be performed every 3 A by a semi-automated procedure called QCOM [39]. This procedure was unique and so successful that currents of many tens of amperes could be safely accumulated. The maximum current recorded in a single ring was 57 A at 26 GeV/*c* and physics beams were typically 30–40 A at 31.4 GeV/*c*.

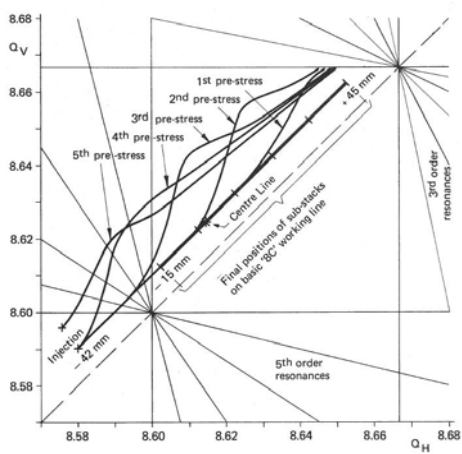

**Fig. 21:** Pre-calculated, pre-stressed working lines

*Impedance and stability criteria*

The ISR was the ideal machine for studies on beam impedance and stability. There were many publications, but some notable topics were the now widely-used longitudinal Keil–Schnell criterion [40], the Schnell–Zotter criterion for the transverse plane [41], and Landau damping.

## 10  Beam–beam interaction

In 1973, Eberhard Keil wrote a short note entitled "Why be afraid of magnet imperfection resonances in high luminosity storage rings?" [42]. He was referring to the beam itself being, in most cases, a stronger source of resonance excitation and one that excites all orders since a Gaussian distribution can be expressed as an infinite power series. The ISR was an ideal and unique test bed for the study of coasting and bunched beams colliding at a small angle. These results were compared at length with those from electron–positron machines and helped to guide the design of future colliders. A large number of papers were published and a useful introduction is given in Ref. [43]. As part of the beam-beam studies the ISR was also equipped with a non–linear lens [44].

## 11  Others topics

The above is far from an exhaustive list. The ISR was also used for the study of intra-beam scattering and overlap knockout resonances. Other examples of special beam equipment are the transverse feedback systems and the scrapers for beam cleaning. Moving more to the physics, there was the development of the Roman pots. Today there is a strong emphasis on technology transfer and the ISR also has examples of this such as the Digital Teslameter designed by Klaus Brand and commercialized by a company in Geneva [45].

## 12  Conclusion

The ISR was a major project with a large experienced staff that had the critical mass to spontaneously generate new ideas and the capacity to follow them up. It attracted visitors, students, fellows, experts on sabbatical leave and many others all of whom helped to catalyse progress. The timing of the project was such that the world-wide community was poised to advance to what we would now recognise as 'modern' accelerators. One could argue that many of the topics described would have occurred whether the ISR was a collider or just a plain synchrotron, but much of what has been described depends rather strongly on the particular attributes of the ISR. Schottky noise exists in a bunched beam, but it is less likely to be discovered. The exceptionally large momentum spread of the ISR pushed the studies of tune shifts, coupling and chromaticity schemes into more detail. The colliding beam geometry was essential for beam–beam studies. Colliding beam physics would certainly have been seriously held back without this full-scale test stand for experimentation. In short, the ISR looked into the part of parameter space that would be needed for the next step in high-energy physics. It was the right machine at the right time. This appears logical now, but it was a brave decision at the time.